\documentstyle[12pt]{article}
\begin{document}

\centerline{\Large\bf The 1999 Heineman Prize Address}

\vspace{.3in}

\centerline{\Large\bf Integrable models in statistical mechanics:}

\vspace{.3in}

\centerline{\Large\bf The hidden field with unsolved problems}

\vspace{.3in}

\centerline{ Barry~M.~McCoy\footnote{e-mail: mccoy@insti.physics.sunysb.edu}}

\vspace{.3in}

\centerline{ Institute for Theoretical Physics}
\centerline{ State University of New York,
 Stony Brook,  NY 11794--3840}

\begin{abstract}

In the past 30 years there have been extensive discoveries in the
theory of integrable statistical mechanical models including the
discovery of non-linear differential equations for Ising model
correlation functions, the theory of random impurities, level
crossing transitions in the chiral Potts model and the use of
Rogers-Ramanujan identities to generalize our concepts of Bose/Fermi
statistics. Each of these advances has led to the further 
discovery of major unsolved problems of great mathematical 
and physical interest. I will here discuss the  mathematical advances, the
physical insights  and extraordinary lack of visibility of this
field of physics.

\end{abstract}

\section{ Introduction}

It is a  great honor to have been awarded the Dannie Heineman Prize in
Mathematical Physics and it is a great pleasure to be be in the
company today of my fellow prize winners, C.N. Yang, this year's winner
of the Lars Onsager prize, A.B. Zamolodchikov, the inventor of conformal
field theory, and especially my collaborator and thesis adviser
T.T. Wu.

\section{Why integrable models is the invisible field of physics}

The citation of this year's Heineman Prize is for ``groundbreaking and
penetrating work on classical
statistical mechanics, integrable models and conformal field theory''
and in this talk I plan to discuss the work in statistical mechanics and
integrable models for which the award was given. But since the
Heineman prize is explicitly for publications in mathematical physics 
I want to begin by examining what
may be meant by the phrase ``mathematical physics.''

The first important aspect of the term ``mathematical physics'' is that
it means something very different from most other kinds of physics.
This is seen very vividly by looking at the list of the divisions of
the APS. Here you will find 1) astrophysics, 2) atomic physics,
3) condensed matter physics, 4) nuclear physics, and 5) particles and
fields but you will not find any division for mathematical
physics. This lack of existence of mathematical physics as a field is
 also reflected
in the index of Physical Review Letters where mathematical physics is
nowhere to be found.
Therefore we see that the Heineman prize in mathematical physics 
is an extremely
curious award because it honors achievements in a field of physics
which the APS does not recognize as a field of physics.

This lack of existence of mathematical physics as a division in the
APS reflects, in my opinion, a deep uneasiness about the relation of
physics to mathematics. An uneasiness I have heard echoed hundreds of
times in my career in the phrase ``It is very nice work but it is not
physics. It is mathematics''. A phrase which is usually used before the
phrase ``therefore we cannot hire your candidate in the physics department.''

So the first lesson to be learned is that mathematical physics is an
invisible field. If you want to
survive in a physics department you must call yourself something else. 

So what can we call the winners of the Heineman prize in mathematical
physics if we cannot call them mathematical physicists? The first
winner was Murray Gell--Mann in 1959. He is surely belongs in particles
and fields; the 1960 winner Aage Bohr is surely a nuclear physicist;
the 1976 winner Stephen Hawking is an astrophysicist. And in fact
almost all winners of the prize in mathematical physics can be classed
in one of the divisions of the APS without much confusion.

But there are at least two past winners who do not neatly fit into
the established categories, Elliott Lieb, the 1978 recipient,
and Rodney Baxter, the 1987 recipient, both of whom have made outstanding
contributions to the study of integrable models in classical
statistical mechanics---the same exact area for which Wu and I are
being honored here today. This field of integrable models in
statistical physics is the one field of mathematical physics which
does not fit into some one of the existing divisions of the APS. 

It is for this reason that I have described integrable  models 
as a hidden field. Indeed it is so hidden that it is sometimes
not even considered it to be
statistical mechanics as defined by the IUPAP.

The obscurity of the field of integrable statistical mechanics 
models explains why there are
less than a dozen physics departments in the United States where it is 
done. This makes the job prospects of a
physicist working in this field very slim. But on the other hand it
means that we in the field get to keep all the good problems to
ourselves. So it is with mixed feelings that I will now proceed to
discuss some of the progress made in the last 33 years and the some of
the directions for future research.

\section{The Ising Model}

The award of the  1999 Heineman prize, even though the citation says that it
is for work on integrable models, is in fact for work done from
1966--1981 on a very specific system: the two dimensional Ising model. 
This work includes 
boundary critical phenomena, randomly layered systems, the Painlev{\' e}
representation of the two point function, and the first explicit results on
the Ising model in a magnetic field. All of these pieces of work had
results which were unexpected at the time and all have lead to
significant extensions of our knowledge of both  statistical mechanics
and mathematics.

\subsection{The boundary Ising model and wetting }

The Ising model is a two dimensional collection of classical ``spins''
$\sigma_{j,k}$ which take on the two values $+1$ and $-1$ and are located
at the $j$ row and $k$ column of a square lattice. For a
translationally invariant system the interaction energy of this system
is
\begin{equation}
{\cal E}=
-\sum_{j,k}(E^v\sigma_{j,k}\sigma_{j+1,k}+E^h\sigma_{j,k}\sigma_{j,k+1}
+H\sigma_{j,k}).
\label{ising}
\end{equation}
This is certainly one of the two most famous and important models in
statistical mechanics (the Heisenberg-Ising chain being the other) and
has been studied by some of the  most distinguished physicists of this
century including Onsager \cite{ons}who computed the free energy for $H=0$
in 1944 and Yang \cite{yanga} who computed 
the spontaneous magnetization in 1952.

In 1966 I began my long involvement with this model when, for part of
my thesis, Prof. Wu suggested that I compute for an Ising model on a
half plane the same quantities which Onsager and Yang computed for the
bulk. At the time we both thought that because the presence of the
boundary breaks translational invariance  the boundary computations
would be at least as difficult as the bulk computations. It was
therefore quite surprising when it turned out that the computations were
drastically simpler \cite{mccoya}. In the first place the model could
be solved in the presence of a field on the boundary which meant that the
computation of the magnetization came along for free once we could do
the free energy but more importantly the correlation functions, which
in the bulk  were given by large determinants whose size increased
with the separation of the spins, were here given by nothing worse than
the product of two one--dimensional integrals for all separations. The
key to this great simplification is the fact that the extra complication
of the boundary magnetic field actually makes the problem simpler to
solve (a realization I had in a dream at  3:00 AM after a New Years
eve party). 

This model is the first case where boundary critical
exponents were explicitly computed. Indeed it remained almost the only
solved problem of a boundary critical phenomena until it was
generalized to integrable massive  boundary  field theory in 1993 \cite{gz}.

This boundary field had the added virtue that we could analytically
continue the boundary magnetization into the metastable region and
explicitly compute a boundary hysteresis effect \cite{mccoyb}. This leads to
the lovely effect that near the value of the boundary magnetic field
where the hysteresis curve ends
that the spins a very long distance from the boundary turn over from
pointing in the direction of the bulk magnetization to pointing in the
direction  of the metastable surface spin. At the value where the
metastability ends this surface effect penetrates all the way to
infinity and ``flips'' the spin in the bulk.  In later years this
phenomena has been interpreted at a ``wetting transition''
\cite{fisher} and the
Ising intuition has been extended to many models where exact
solutions do not exist.

\subsection{Random layered Ising model and 
Griffiths-McCoy singularities}

Our next major project was to generalize the translationally invariant
interaction (\ref{ising}) to a non translationally invariant problem where
not just a half plane boundary was present but to a case where
\begin{enumerate}
\item The interaction energies $E^v$ were allowed to vary from row to
row but translational invariance in the horizontal direction was
preserved
\item The interaction energies $E^v(j)$ between the rows were chosen
as independent random variables with a probability distribution $P(E^v).$
\end{enumerate}

This was the first time that such a random impurity problem had ever
been studied for a system with a phase transition and the entire
computation was a new invention \cite{mccoyc}. 
In particular we made the first use in
physics of Furstenburg's theory \cite{furs} 
of strong limit theorems for matrices.
We felt that the computation was a startling success because we found
that for any probability distribution, no matter how small the variance, 
there was a temperature scale, depending on the variance, where there
was new physics that is not present in the translationally invariant
model.  For example the divergence in the specific heat at $T_c$ decreases from
the logarithm of Onsager to an infinitely differentiable essential singularity.
Moreover the average over the distribution $P(E^v)$ of the correlation 
functions of the boundary could be computed \cite{mccoyd} and it 
was seen that there
was an entire temperature range surrounding $T_c$ where the  boundary
susceptibility was infinite. Thus the entire picture of critical exponents
which had been invented several years before to describe critical
phenomena in pure systems was not sufficient to describe these random
systems \cite{mccoye}. We were very excited.

But then something happened which I found very strange. Instead of attempting  
to further explore the physics we had found, arguments were given as to
why our effect could not possibly be relevant to real systems. This has
lead to arguments which continue to this day.

We were, and are, of the opinion that the effects seen in the layered
Ising model are caused by the fact that the zeroes of the partition
function do not pinch the real temperature axis at just one point but
rather pinch in an entire line segment. A closely related effect
was simultaneously discovered by Griffiths \cite{griffiths} and it was, and is, our
contention that in this line segment there is a new phase of the
system. But this line segment is not revealed by approximate
computations and for decades it was claimed that our new phase was
limited to layered systems; a claim which some continue to make to
this day.

However, fortunately for us, there is an alternative interpretation of
the Ising model in terms of a one--dimensional quantum spin chain in
which our layered classical two--dimensional system becomes a randomly
impure quantum chain \cite{sm}. In this interpretation 
there is no way to argue  
 away the existence of our new phase and finally in 1995, a quarter
century after we first found the effect, D. Fisher \cite{dfisher}, in an astounding
paper, was able to craft a theory of the physics of rare events based
on an exact renormalization group computation which
not only reproduced our the results of our layered model on the
boundary but  extended the computations to bulk quantities which in
1969 we had been unable to compute. With this computation I think 
 the existence of what are now called Griffiths-McCoy singularities 
is accepted, but it has taken a quarter century for this to happen.

\subsection{Painlev{\' e} Functions and difference equations}

But perhaps the most dramatic discoveries were  
 published from 1973 to 1981 on the  
spin correlation functions of the Ising model and most particularly the
results that in the scaling limit where $T\rightarrow T_c$ and the
separation of the spins $N \rightarrow \infty$ such that $|T-T_c|N=r$
is fixed that the correlation function \cite{mccoyf}
$<\sigma_{0,0}\sigma_{0,N}>$
divided by $|T-T_c|^{1/4}$ is
\begin{equation}
G_{\pm}(2r)=(1\mp\eta)\eta^{-1/2}{\rm exp}\int_r^{\infty}dx{1\over
4}x\eta^{-2}[(1-\eta^2)^2-\eta'^2]
\label{scaling}
\end{equation}
where $\eta(x)$ satisfies the third equation of  Painlev{\' e} \cite{pain}
\begin{equation}
{d^2\eta\over dx^2}={1\over \eta}({d\eta\over dx})^2-{1\over
x}{d\eta\over dx}+{1\over
x}(\alpha\eta^2+\beta)+\gamma\eta^3+{\delta\over \eta}
\label{pain}
\end{equation}
with $\alpha=\beta=0$ and $\gamma=-\delta=1$ and 
\begin{equation}
\eta(x)\sim 1-{2\over \pi}K_0(2x)~~~{\rm as}~~x\rightarrow \infty
\end{equation} 
where $K_0(2x)$ is the modified Bessel function of order zero.
Furthermore on the lattice all the correlation functions satisfy
quadratic difference equations \cite{mccoyg}.

This discovery of Painlev{\' e} equations in the Ising model was the
beginning of a host of developments in mathematical physics which
continues in an ever expanding volume to this day. It led Sato, Miwa,
and Jimbo \cite{smj} to their celebrated series of work on isomonodromic
deformation and to the solution of the distribution of eigenvalue of
the GUE random matrix problem \cite{jmms} in terms of a Painlev{\'e} V
function. This has subsequently been extended by many people, including
one of our original collaborators, Craig Tracy \cite{tw}, 
to so many branches of
physics and mathematics including random matrix theory matrix models
in quantum gravity and random permutations that entire semester long
workshops are now devoted to the subject. Indeed a recent book on
special functions \cite{iksy} characterized 
Painlev{\' e} functions as ''the special
functions of the 21st century.'' Rarely has the solution to one
small problem in physics had so many ramifications in so many
different fields.

\subsection{ Ising model in a field}

The final piece of work in the Ising model to be mentioned is 
what happens to the two point function  when a 
small magnetic field is put the system for $T<T_c.$ At $H=0$ for
$T<T_c$ the two point function has the important property that it
couples only to states with an even number of particles and thus, in
particular the leading singularity in the Fourier transform is not a
single particle pole but rather a two particle cut. 
In 1978 \cite{mccoyh}, as an
application of our explicit formulas for the $n$ spin
correlation functions \cite{nspin} we did a (singular) perturbation computation
to see what happens when in terms of the scaled variable $h=H/|T-T_c|^{15/8}$
a small value of $h$ is applied to the system. We found that the two
particle cut breaks up into an infinite number of poles which are 
given by the zeroes of Airy functions. These poles are exactly at the
positions of the bound states of a linear potential and are immediately
interpretable as a weak confinement of the particles which are free at
$H=0.$ This is perhaps the earliest explicit computation where 
confinement is  seen. From this result it was natural to conjecture
that as we take the Ising model from $T>T_c,~H=0$ to $T<T_c,~H=0$ that
as $h$ increases from $0$ to $\infty~(T=T_c,H>0)$ that bound states
emerge from the two particle cut and that as we further proceed from
$T=T_c,~H>0$ down to $T<T_c,~H=0$ that bound states continue to emerge
until at $H=0$ an infinite number of bound states have emerged and
formed a two particle cut. What this picture does not indicate is the
remarkable result found 10 years later by A. Zamolodchikov \cite{zamb}
that at $T=T_c,~
H=0$ the problem can again be studied exactly. This totally unexpected
result will be discussed by Zamolodchikov in the next presentation.

\section{From Ising to integrable}

Even at the time when this Ising model work was 
initiated there were other models known such as
the Heisenberg chain \cite{bethe}, \cite{yy}
the delta function gases \cite{lieba}-\cite{yangb}, the 6--vertex
model \cite{liebb},\cite{yangc}, 
the Hubbard model \cite{liebc}, and the 8--vertex model \cite{bax} 
for which the
free energy (or ground state energy), and 
the excitation spectrum could be computed exactly. Since then it has 
been realized that a fundamental equation first seen by
Onsager \cite{ons},\cite{wannier} in the Ising model 
and used in a profound way by Yang 
\cite{yangb} in the
delta function gases and by Baxter \cite{bax} in the 8 vertex model could be
used extend these computations to find many large classes of 
models for which free energies
could be computed. These models which come from the Yang-Baxter (or
star triangle equation) are what are now called the integrable models.

The Ising model itself is the simplest case of such
an integrable model. It thus seems to be a very natural conjecture
which was made by Wu, myself and our collaborators the instant we
made the discovery of the Painlev{\' e} representation of the Ising
correlation function that there must be a similar representation for the
correlation functions of all integrable models. To be more precise I
mean the following

{\bf Conjecture}

{\it The correlation functions of integrable statistical mechanical models
are characterized as the solutions of classically integrable equations
(be they differential, integral or difference).}

One major step in the advancement of this program was made by our next
speaker, Alexander
Zamolodchikov, who showed, with the invention of
conformal field theory \cite{bpz}, that this conjecture 
is realized for models at the
critical point. One of the major unsolved problems of integrable models
today is to extend the linear equations which characterize correlation
functions in conformal field theory to
nonlinear equations for massive models. This will realize the goal of
generalizing to all integrable models what we have learned for the
correlation functions of the Ising model. This is an immense field of
undertaking in which many people have and are making major
contributions. It is surely not possible to come close to surveying
this work in the few minutes left to me. I will therefore confine myself
to a few remarks about things I have been personally involved with
since completing the work with Wu in 1981 on the Ising model.

\subsection{The chiral Potts model}

In 1987 my coauthors H. Au-Yang, J.H.H. Perk, C.H. Sah, S. Tang, and M.L.
Yan and I discovered \cite{ampty} the first example of 
an integrable model where, in
technical language, the spectral
variable lies on a curve of genus higher than one. This model has $N
\geq 3$
states per site and is known
as the integrable chiral Potts model. It is a particular case of a
phenomenological model introduced for case $N=3$
in 1983 by Howes, Kadanoff and den
Nijs \cite{hkd} in their famous study of level crossing transitions and
is a generalization of the $N$ state model introduced by Von Gehlen
and Rittenberg in 1985 \cite{gr} which generalizes Onsager's original
solution of the Ising model \cite{ons},\cite{gst}.

The Boltzmann weights were
subsequently shown by Baxter, Perk and Au-Yang \cite{bpa} to have the following
elegant form for $0\leq n \leq N-1$
\begin{equation}
{W^h_{p,q}(n)\over
 W^h_{p,q}(0)}=\prod_{j=1}^{n}({d_pb_q-a_pc_q\omega^j
\over b_pd_q-c_pa_q\omega^j}),~~
 {W^v_{p,q}(n)\over W^v_{p,q}(0)}=\prod_{j=1}^n({\omega
 a_pd_q-d_pa_q\omega^j\over c_pb_q-b_pc_q\omega^j})
\label{cpw}
\end{equation}
where  $\omega = e^{2 \pi i/N}.$ The variables $a_p,b_p,c_p,d_p$ and
$a_q,b_q,c_q,d_q$ satisfy the equations
\begin{equation}
a^N+kb^N=k'd^N,~~~ka^N+b^N=k'c^N
\label{curve}
\end{equation}
with $k^2+k'^2=1$ and this specifies a curve of genus $N^3-2N^2+1.$
When $N=2$ the Boltzmann weights reduce to those of the Ising model
(\ref{ising}) with $H=0$ and the curve (\ref{curve}) reduces to the
elliptic curve of genus 1. However when $N\geq 3$ the curve has genus
higher than one. This is the first time that such higher genus curves
has arisen in the Boltzmann weights of integrable models. 

This model is out of the class of all previously known models and
raises a host of unsolved questions which are related to some of the
most intractable problems of algebraic geometry which have been with
us for 150 years. As an example of these new occurrences of ancient
problems we can consider the spectrum of the  transfer matrix 
\begin{equation}
T_{\{l,l'\}}=\prod_{j=1}^{\cal N}W_{p,q}^v(l_j-l'_j)W_{p,q}^h(l_j-l'_{j+1}).
\end{equation}
This transfer matrix satisfies the commutation relation
\begin{equation}
[T(p,q),T(p,q')]=0
\end{equation}
and also satisfies functional equations on the Riemann surface at
points connected by the automorphism 
$R(a_q,b_q,c_q,d_q)=(b_q,\omega_q,d_q,c_q).$ For the Ising case $N=2$
this functional equation reduces to an equation which can be solved
using elliptic theta functions. Most unhappily, however, for the
higher genus case the analogous solution requires machinery from
algebraic geometry which does not exist. For the problem of the free
energy Baxter \cite{baxfree} has devised an ingenious method of solution which
bypasses algebraic geometry completely but even here some problems
remain in extending the method to the complete eigenvalue spectrum \cite{mr}.

The problem is even more acute for the order parameter of the
model. For the $N$ state models there are several order parameters
parameterized by an integer index $n$ where $1\leq n\leq N-1.$ For
these order parameters $M_n$ we conjectured \cite{al} 
10 years ago from 
perturbation theory computations that 
\begin{equation}
M_n=(1-k^2)^{n(N-n)/2N^2}.
\end{equation} 
When $N=2$ this is exactly the result announced by Onsager
\cite{onsb}  in 1948 and 
proven by Yang \cite{yanga} for the Ising model in 1952. For the Ising model it
took only three years to go from conjecture to proof. But for the
chiral Potts model a decade has passed and even though Baxter 
\cite{baxe}--\cite{baxc} has
produced several elegant formulations of the problem which all lead to
the correct answer for the Ising case none of them contains enough
information to solve the problem for $N\geq 3.$ In one approach
\cite{baxe} the
problem is reduced the the evaluation of a path ordered exponential of
noncommuting variables on a Riemann surface. This sounds exactly like
problems encountered in non Abelian gauge theory but, unfortunately,
there is nothing in the field theory literature that helps. In another
approach \cite{baxc} a major step in the solution involves the explicit
reconstruction of a meromorphic function from a knowledge of its
zeros and poles. This is a classic problem in algebraic geometry for
which in fact no explicit answer is known either. Indeed the unsolved 
problems arising from the chiral Potts model are so resistant to all
known mathematics that I have reduced my frustration to the following
epigram;

{\it The nineteenth century saw many brilliant creations of the human
mind. Among them are algebraic geometry and Marxism. In the late
twentieth century Marxism has been shown to be incapable of solving any
practical problem but we still do not know about algebraic geometry.}

It  must be stressed again that the chiral Potts model was not
invented because it was integrable but was found to be integrable
after it was introduced to explain experimental data. In a very
profound way physics is here far ahead of mathematics.

\subsection{Exclusion statistics and Rogers-Ramanujan identities}

One particularly important property of integrable systems is seen in
the spectrum of excitations above the ground state. In all known cases
these spectra are of the quasiparticle  form in which the energies of 
multiparticle states are additively composed of single particle
energies $e_{\alpha}(P)$ 
\begin{equation}
E_{ex}-E_0=\sum_{\alpha=1}^n\sum_{j=1}^{m_\alpha}e_{\alpha}(P_j^{\alpha})
\end{equation}
with the total momentum
\begin{equation}
P=\sum_{\alpha}^n\sum_{j=1}^{m_\alpha}P^{\alpha}_j~~({\rm mod}~2\pi).
\end{equation}
Here  $n$ is the number of types of quasi-particles and there are
$m_{\alpha}$ quasiparticles of type $\alpha.$
The momenta in the allowed states are 
quantized in
units of $2\pi/M$ and are chosen from the sets
\begin{equation}
P^{\alpha}_j\in \{P_{\rm min}^{\alpha}({\bf m}),
P_{\rm min}^{\alpha}({\bf m})+{2\pi\over M},
P_{\rm min}^{\alpha}({\bf m})+{4\pi\over M},\cdots,
P_{\rm max}^{\alpha}({\bf m})\}
\label{rules}
\end{equation}
with the Fermi exclusion rule 
\begin{equation}
P_{j}^{\alpha}\neq P_{k}^{\alpha}~~{\rm for}
j\neq k~~{\rm  and~ all}~~ \alpha
\label{fermi}
\end{equation}
and
\begin{equation}
P^{\alpha}_{\rm min}({\bf m})={\pi\over M}[({\bf m}({\bf
B}-1))_{\alpha}-A_{\alpha}+1]~~{\rm and}~~  
P^{\alpha}_{\rm max}=-P^{\alpha}_{\rm min} +{2\pi\over M}({{\bf
u}\over 2}-{\bf A})_{\alpha}
\end{equation}
where if some $u_{\alpha}=\infty$ the corresponding
$P_{\rm max}^{\alpha}=\infty.$ 

If some $e_{\alpha}(P)$ vanishes at some momentum (say 0) the system
is massless and for $P\sim 0$ a typical behavior is $e_{\alpha}=v|P|$
where $v$ is variously called the speed of light or sound or the
spin wave velocity.

The important feature of the momentum selection rules (\ref{rules}) is
that in addition to the fermionic exclusion rule (\ref{fermi}) is the
exclusion of a certain number of momenta at the edge of the momentum zones
which is proportional to the number of quasiparticles in the state. 
For the Ising model at zero field there is only one quasiparticle 
and $P_{\rm min}=0$ so the quasiparticle is exactly the same as a free
fermion. However, for all other cases the $P_{\rm min}$ is not zero
and exclusion does indeed take place. This is a very explicit
characterization of the generalization which general integrable models
make over the Ising model. 

The exclusion rules (\ref{rules}) lead to what have been called fractional 
(or exclusion) statistics by Haldane \cite{hal}.
On the other hand they make a remarkable and beautiful
connection with the mathematical theory of Rogers-Ramanujan 
identities and conformal field theory. We have found that these
exclusion rules allow a reinterpretation of all conformal field
theories (which are usually discussed in terms of a bosonic Fock space
using a Feigin and Fuchs construction \cite{ff}) in terms of a
set of fermionic quasiparticles \cite{kkmm}. What is most surprising
is that there is not just one fermionic representation for each
conformal field theory but there are at least as many distinct
fermionic representations as there are integrable perturbations. The
search for the complete set of fermionic representations is ongoing
and I will only mention here that we have extensive results for the
integrable perturbations of the minimal models $M(p,p')$ for the
$\phi_{1,3}$ perturbation \cite{bms} and the $\phi_{2,1}$ and
$\phi_{1,5}$ perturbations \cite{bmp}.

\section{Beyond Integrability}

There is one final problem of the hidden field of integrable models
which I want to discuss. Namely the question of what is the relation
of an integrable model to a generic physical system which does not
satisfy a Yang-Baxter equation. For much of my career I have been
told by many that these models are just mathematical curiosities which
because they are integrable can, by that very fact, have nothing to do
with real physics. But on the other hand the fact remains that all of
the phenomenological insight we have into real physics phase
transitions as embodied in the notions of critical exponents,
scaling theory and universality which have served us well for 35 years
either all come from integrable models or are all confirmed by the
the solutions of integrable models. So if integrable models leave
something out we have a very poor idea of what it is.

Therefore it is greatly interesting that several months ago Bernie
Nickel \cite{bn}
sent around a preprint in which he made the most serious advance in
the study of the Ising model susceptibility since our 1976 paper
\cite{mccoyf}. In that paper in addition to the
Painlev{\' e} representation of the two point function we derive  an
infinite series for the Ising model susceptibility where the $n^{th}$ term
in the series involves an $n^{th}$ order integral. 

When the integrals in this expansion are
scaled to the critical point each term contributes to the leading
singularity of the susceptibility $|T-T_c|^{-7/4}$ However Nickel goes
far beyond this scaling and for the isotropic case where $E^v=E^h=E$ 
in term of the variable $v=\sinh  2E/kT$ he shows that successive
terms in the series contribute singularities that eventually become
dense on the unit circle in the complex plane $|v|=1.$ From this he
concludes that unless unexpected cancelations happen that there will
be natural boundaries  in the susceptibility on $|v|=1$. This would
indeed be a new effect which could make integrable models different
from generic models, Such natural boundaries have been suggested by
several authors in the past including Guttmann \cite{gut}, and Orrick
and myself \cite{mo} on the basis of perturbation studies 
of nonintegrable models 
which show ever increasingly complicated singularity structures as the
order of perturbation increases; a complexity which magically
disappears when an integrability condition is  imposed. This connection
between integrability and analyticity was first emphasized 
by Baxter \cite{baxd}
long ago in 1980 when he emphasized that the Ising model in a magnetic
field satisfies a functional equation very analogous to the zero field
Ising model but that the Ising model in a field lacks the analyticity
properties need for a solution. The proof of Nickel's
conjecture will, if correct, open up a new view on what it means to be
integrable.

\section{Conclusion}

I hope that I have conveyed to to you some of the excitement and
challenges of the field of integrable models in statistical mechanics.
The problems are physically important, experimentally accessible, and
mathematically challenging. The field has been making constant
progress since the first work of Bethe in 1931 and Onsager in
1944. So it might be thought that, even though the problems are
hard,  it would command the attention of some of the most powerful
researchers in a large number of institutions. But as I indicated in
the beginning of this talk this is in fact not the case. 

Most physics departments are more or less divided into the same
divisions as is the APS. Thus it is quite typical to find departments
with a condensed matter group, a nuclear physics group, and high energy
group, an astrophysics group and an atomic and molecular group. But as
I mentioned at the beginning, none of the work I have discussed in this
talk fits naturally into these categories and thus if departments hire
people in the mainstream of the existing divisions of the APS no one
doing research in integrable models in statistical mechanics will ever
be hired. 

So while I am deeply honored and grateful for the award of the 1999
Heineman prize for mathematical physics there is still another honor I
am looking for. It is to receive a letter
from the chairman of a physics department which reads as follows:

{\it \noindent {Dear Prof. McCoy,}

Thank you for the recommendation you recently made to us concerning
the hiring of a new faculty member. We had not considered hiring
anyone in the area of physics represented by your candidate, but after
reading the resume and publications we decided that you were
completely correct that the candidate is doing outstanding work
which will bring  an entirely new area of research to our department. We
are very pleased to let you know that the university has made
an offer of a faculty appointment to your candidate which has been
accepted today. Thank you very much for your help and advice.}  

I have actually received one such letter in my life. If I am fortunate
I hope to receive a few more before the end of my career.
The 21st century is long and anything is still possible.

\vspace{.5in}

{\bf Acknowledgments}

This work is supported in part by the National Science Foundation
under grant DMR 97-03543.

\end{document}